\documentclass[11pt]{extarticle}

\usepackage{graphicx}
\usepackage{amssymb}
\usepackage{bm}
\usepackage{amsmath}
\usepackage{algorithmic}
\usepackage{authblk}
\usepackage{hyperref}
\usepackage{bbm}
\usepackage{listings}

\def\x{\textit{\textbf{x}}}
\def\y{\textit{\textbf{y}}}
\def\v{\textit{\textbf{v}}}

\def\n{\textit{\textbf{n}}}
\def\p{\textit{\textbf{p}}}

\def\R{{\mathbb{R}}}
\def\dive{\mathop{\mathrm{div}}}
\def\btheta{{\mbox{\boldmath$\theta$}}}

\begin{document}

\title{PyFRep: Shape Modeling with Differentiable Function Representation}

\author{Pierre-Alain Fayolle\footnote{University of Aizu, \href{mailto:fayolle@u-aizu.ac.jp}{fayolle@u-aizu.ac.jp}} 
\and Evgenii Maltsev\footnote{Skoltech, \href{mailto:evgenii.maltsev@gmail.com}{evgenii.maltsev@gmail.com}}
}
\date{}

\maketitle

\begin{abstract}
\small
We propose a framework for performing differentiable geometric modeling based on the Function Representation (FRep). The framework is built on top of modern libraries for performing automatic differentiation allowing us to obtain derivatives w.r.t. space or shape parameters. We demonstrate possible applications of this framework: Curvature estimation for shape interrogation, signed distance function computation and approximation and fitting shape parameters of a parametric model to data. Our framework is released as open-source \footnote{\url{https://github.com/fayolle/PyFRep}\label{repo_url}}.
\end{abstract}

\section{Introduction} 
The ability to compute derivatives via automatic differentiation is fundamental for performing backpropagation, which is the workhorse behind deep learning. It is now starting to play an important role in areas such as scientific computing, computer graphics, or CAD systems. 

For example, differentiable rendering allows us to solve tasks of inverse rendering, which consists in restoring information about the scene from an image, such as the camera position, the intensity and position of the light source, model parameters, etc. Another example is gradient-based optimization, which is widely used in image restoration, image registration, stereo vision, optical flow estimation, and others. In CAD-based shape optimization and numerical optimization tasks, geometric sensitivities (gradients of CAD surface points w.r.t. design parameters of the model) are important and also require an efficient and precise computation of derivatives. Finally, information about the derivatives can also be useful in 3D modeling, rendering, and meshing algorithms.

In this paper, we describe a framework for differentiable geometric modeling based on the Function Representation (FRep) \cite{pasko1995function} and show possible applications. FRep allows one to define a geometric shape by means of a continuous real function. We add as a requirement that the function is differentiable almost everywhere. We consider possible applications of this framework, such as the evaluation of curvature information, which is useful for shape interrogation, signed distance function computation and approximation, and fitting the parameters of a model to data. Our framework is implemented in Python and is released as open-source \footref{repo_url}.

\section{Related Work}
\paragraph{Programmatic shape modeling systems}
Multiple programmatic shape modeling systems have been proposed over the years. This includes systems based on implicit surface modeling, such as the BlobTree \cite{Wyvill99,Schmidt07}, and systems based on the Function Representation (FRep) including HyperFun \cite{pasko1999hyperfun}, variants targeting the Java-Virtual Machine \cite{cartwright2005web,fayolle2005web}, or more recent implementations such as LibFive \cite{keeterLibFive,keeter2020sig}. 

OpenSCAD \cite{openscad} is a popular programmatic system, based on representing solids by the Boundary Representation. It seems to be especially popular in the rapid prototyping community. 

Another popular system in the rapid prototyping community is IceSL \cite{lefebvre2013icesl}. It is a hybrid system that relies on implicit surfaces, signed distance functions, and polygon meshes. A similar system is OpenFab \cite{vidimce2013} with an emphasis on modeling multimaterial objects. 

Similarly to these systems, we propose a programmatic shape modeling system. However, we build it on top of automatic differentiation libraries so that our shape models are differentiable with respect to spatial coordinates or shape parameters. We use these facilities for shape queries or to fit parametric models to datasets. 

\paragraph{Automatic differentiation}
Automatic differentiation refers to techniques used to obtain the derivatives of a function defined as a computer program (a computational graph). These techniques automate derivative computation by applying the chain rule. There are two different modes for automatic differentiation: Forward mode and reverse mode. Each mode corresponds to the order in which the chain rule is applied while traversing the computational graph. Accumulating derivatives from the left (the input) corresponds to the forward mode, while accumulating derivatives from the right (the output) corresponds to the reverse mode. The reverse mode is more efficient for scalar-based functions, while the forward mode is more efficient for vector-based functions and when higher-order derivatives are necessary. The reader can refer to the book by Griewank and Walther \cite{griewank2008} or the recent survey \cite{baydin2017} for further details. 

Automatic differentiation is efficient since it allows to compute the gradient with the same (asymptotic) time complexity as the function evaluation unlike, for example, finite differences. 

There is a growing interest in automatic differentiation as part of deep learning frameworks for the implementation of the backpropagation algorithm \cite{pearlmutter2008reverse,maclaurin2016modeling} which is used to train neural networks. Popular frameworks include PyTorch \cite{paszke2019pytorch}, Tensorflow \cite{tensorflow2015}, or JAX \cite{jax2018github} among others. 

\paragraph{Differentiable systems in graphics}
Multiple attempts have been made in the computer graphics and computer vision community to develop differentiable rendering systems. A differentiable renderer supplies the derivatives of the image pixels w.r.t. the scene parameters (camera parameters, models, \ldots)

One of the first differentiable rendering systems, OpenDR \cite{loper2014opendr}, was applied to the fitting of a human body to images and depth images. Several other systems have been developed subsequently for differentiable rendering, such as \cite{anderson2017aether,li2018montecarlo,li2018halide,li2020differentiable}. Differentiable renderers for shader design were recently proposed in \cite{guo2020bayesian,shi2020match}. Other differentiable systems were proposed for optimization in image processing and graphics \cite{devito2017opt} or physical simulation \cite{hu2019difftaichi}. For recent surveys on the topic, see \cite{kato2020differentiable,tewari2020survey}. 

These systems deal with differentiable rendering algorithms. We deal with a system for programmatically expressing a shape and its derivatives, where the derivatives are computed w.r.t. spatial or shape parameters. With the current advances in generative AI, we believe that programmatic shape modeling systems can play an important role. See, for example, the discussion in \cite[Section~5]{Fayolle_csg2024}. 

Closer to this work is the approach described in \cite{mykhaskiv2018nurbs} which is based on NURBS, while our system is based on the Function Representation. The use of differentiable geometric kernels is useful for the simulation and optimization of CAD models \cite{hafner2019x}.

\section{Background}
The goal of this work is to describe a differential geometric kernel based on the Function Representation \cite{pasko1995function}. In this section, we provide background on Function Representation-based modeling and on automatic differentiation. 

\subsection{Function Representation}
\label{sec:background_frep}
The Function Representation \cite{pasko1995function} is an approach to represent a solid by a scalar function $f: \R^3 \rightarrow \R$. The surface of the solid corresponds to the set $\{\x\in\R^3: f(\x)=0\}$, the interior to $\{\x\in\R^3: f(\x)>0\}$ and the exterior to $\{\x\in\R^3: f(\x)<0\}$. Complex solid objects are built by considering \emph{primitive} objects that are combined with \emph{operations} \cite{ricci1973,bloomenthal1990,pasko1995function,gomes2009}. 

Modeling solids with implicit surfaces or with signed distance functions is becoming increasingly popular, as it allows, for example, the representation of tiled and repetitive patterns, the so-called microstructures, that are useful with 3D printing technologies in biomedical applications and other industries \cite{pasko2011procedural}. 

\paragraph{Primitives}
Examples of primitives include simple objects such as spheres, cylinders, cones, cuboids, skeleton-based primitives, polynomials or objects obtained by fitting splines to point-clouds. 

\paragraph{Operations}
The operations applied to the primitives include affine transformations, such as translation, rotation, scaling, and shearing. The representation makes it easy to define Constructive Solid Geometry (CSG) operations using R-functions \cite{shapiro2007semi,pasko1995function} 
\begin{itemize}
    \item Union $S_1 \cup S_2 := f_{S_1} + f_{S_2} + \sqrt{f_{S_1}^2 + f_{S_2}^2}$, 
    \item Intersection $S_1 \cap S_2 := f_{S_1} + f_{S_2} - \sqrt{f_{S_1}^2 + f_{S_2}^2}$, 
    \item Complement $\overline{S} := -f_S$, 
    \item Difference $S_1 \setminus S_2 := S_1 \cap \overline{S_2}$, 
\end{itemize}
where $S_i$ is the solid corresponding to the set $\{x \in \mathbb{R}^3: f_{S_i} \geq 0\}$ ($i=1,2$). Alternative implementations for Boolean operations include $\min/\max$ and their variants. It is also easy to define operations such as blending, sweeping, offsetting/shelling, morphing, and others \cite{pasko1995function}. 

\paragraph{SDF based modeling}
A popular subset of the Function Representation is obtained by working only with Signed Distance Functions (SDF). A function $f$ is an SDF if $f(\x)$ corresponds to the signed (Euclidean) distance from $\x \in \R^3$ to a given surface $\partial S$. The sign is used to determine whether the point $\x$ is inside the solid bounded by the surface $\partial S$ or outside. 

SDF can be built with a constructive approach using normalized CSG operations \cite{biswas2004approximate,fayolle2005sardf} or with a numerical procedure. They have applications in material modeling \cite{biswas2004heterogeneous} or computational physics \cite{osher2006level} among others.

\subsection{Automatic Differentiation}
We consider a function $f: \R^d \rightarrow \R$ for which we want to compute the derivative. We assume that the function is built by composition of a set of primitives with known derivatives. Note that these primitives can eventually be vector-valued (i.e. mappings $\R^M \rightarrow \R^N$). Computing the derivative of $f$ is done by repeatedly applying the chain rule. 

Assume that for $f = f_3 \circ f_2 \circ f_1$ we have
$$
y = f(\x) = f_3(f_2(f_1(\x))), \qquad \x \in \R^d, \quad y \in \R.
$$
Let us note
$$
\x_1 = f_1(\x), \qquad \x_2 = f_2(\x_1).
$$
The gradient (or Jacobian) of $f$ is given by
\begin{equation}\label{f'}
f^{\prime}(\x) = \frac{\partial y}{\partial \x} 
= \frac{\partial y}{\partial \x_2} 
\frac{\partial \x_2}{\partial \x_1}
\frac{\partial \x_1}{\partial \x}, 
\end{equation}
and 
$$
f_3^{\prime}(\x_2) = \frac{\partial y}{\partial \x_2}, \quad
f_2^{\prime}(\x_1) = \frac{\partial \x_2}{\partial \x_1}, \quad 
f_1{\prime}(\x) = \frac{\partial \x_1}{\partial \x}, 
$$
where $f_1^{\prime}, f_2^{\prime}$ and $f_3^{\prime}$ are the Jacobians of the primitives $f_1, f_2$ and $f_3$. 

Each term in (\ref{f'}) is a matrix. The forward mode in automatic differentiation corresponds to performing the matrix multiplications from right to left, while the reverse mode corresponds to performing the multiplications from left to right. In the case of shape modeling with implicits, the function $f$ is scalar-valued, thus reverse mode is more efficient. 

In terms of implementation, a composite function $f$ corresponds to a directed acyclic graph of primitive functions. Evaluating the function corresponds to a forward traversal of the graph. In order to implement the reverse mode, one has to accumulate intermediate values during the forward pass. The gradient is obtained by traversing the graph in reverse and computing $\frac{\partial y}{\partial \x_i}$ for the intermediate values $\x_i$. 

\section{The PyFRep framework}
\label{sec:system}
We represent a solid by a scalar function $f: \R^3 \rightarrow \R$, differentiable almost everywhere. The surface of the solid corresponds to the zero level-set of $f$, i.e. the set $\{\x \in \R^3: f(\x)=0\}$. We use $f>0$ for the interior of the solid as a convention. 

The term differentiable almost everywhere is used to indicate that the function is differentiable everywhere except on a set of measure zero. For example, the function $\min(x,y)$ is not differentiable only when $x=y$. It is interesting to note that PyTorch \cite{paszke2019pytorch} uses, for example, the sub-gradient $(0,1)$ when $x=y$ for its implementation of functions $\min$ and $\max$. In practice, one has only to be careful in preventing the propagation of NaN values. Using sub-gradients or sub-differentials is a possible strategy. Another possibility is to consider $C^1$, or higher-order, variants of the primitives and the operations when available, such as, for example, (5) in \cite{pasko1995function}. In practice, the lack of differentiability at some points did not pose any problem based on our experiments and in the applications that we considered. 

The function $f$ corresponding to a solid is represented by a computer program. A subset of Python is used as the programming language. The derivatives of $f$ are computed by automatic differentiation using reverse mode. We rely on PyTorch \cite{paszke2019pytorch} in the current implementation, though any other autodiff library \cite{tensorflow2015,jax2018github} could be used as a backend. 

\begin{figure}[hbtp]
\begin{center}
\begin{minipage}{0.7\textwidth}
\begin{lstlisting}[language=Python,basicstyle=\small]
def model(x):
    sp1 = sphere(x,center=(0,0,0),r=1)
    b1 = block(x,
               vertex=(-0.75,-0.75,-0.75), 
               dx=1.5,dy=1.5,dz=1.5)
    t1 = intersection(sp1,b1)
    c1 = cylX(x,center=(0,0,0),r=0.5)
    c2 = cylY(x,center=(0,0,0),r=0.5)
    c3 = cylZ(x,center=(0,0,0),r=0.5)
    t2 = difference(t1,c1)
    t3 = difference(t2,c2)
    t4 = difference(t3,c3)
    return t4
\end{lstlisting}
\end{minipage}
\hspace{-1cm}
\begin{minipage}{0.35\textwidth}
 \vspace{3.1cm}
 \includegraphics[width=0.95\textwidth]{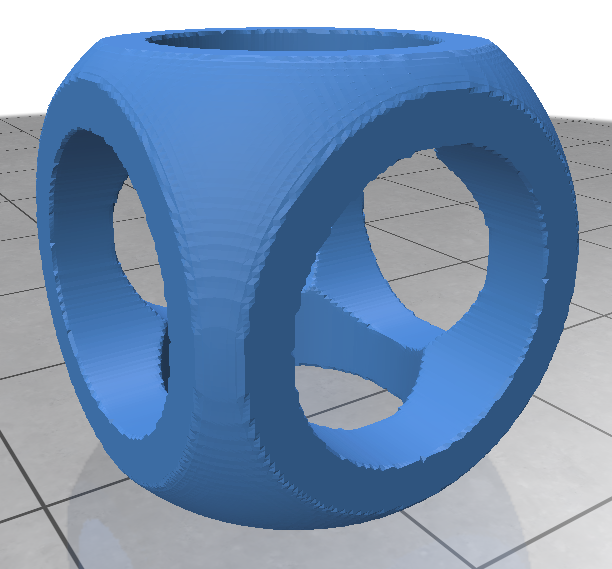}
\end{minipage}
\end{center}
\caption{Example of a program defining a simple shape (left) and the corresponding solid (right).}\label{fig:simple_object}
\end{figure}

Usually, solid objects are not built from scratch but are defined by applying geometric operations to simpler primitives, see Section~\ref{sec:background_frep}. We have implemented all the primitives and operations typically found in implicit-based modeling packages. An example of a simple solid and the corresponding function are given in Fig.\,\ref{fig:simple_object}. The corresponding solid is shown on the right. 

All primitives and operations are implemented in pure Python and are decoupled from the part performing the computation of the derivatives. A computational graph, tracking derivatives computed by reverse mode, is obtained by calling a Python program $f(.)$, representing a solid object, with a variable $\x$ set to require derivatives computation. The derivatives are then obtained by reverse mode automatic differentiation. This approach makes it easy to use the system with different backends such as, for example, PyTorch \cite{paszke2019pytorch}, Tensorflow \cite{tensorflow2015} or JAX \cite{jax2018github}, since the core of primitives and operations is written in pure Python. 

\paragraph{Signed Distance Function}
A popular subset of FRep is the set of Signed Distance Functions (SDF). A function $f$ is an SDF if $f(\x)$ gives the signed (Euclidean) distance from $\x\in\R^3$ to a given surface $\partial S$. We implemented all the usual SDF primitives and operations \cite{fayolle2005sardf,quilez2008modeling}. Currently, SDF primitives and operations, and non-SDF primitives and operations live in two separate modules. Of course, it is possible to model solids by mixing SDF and non-SDF primitives and operations, but the distance property of the overall function is lost. Thus, if $f$ is a function built from a mixture of SDF operations and primitives, and non-SDF operations and primitives, then $f(\x)$ no longer corresponds to the Euclidean distance from the point $\x$ to the surface $\partial S$. SDF allows for a more efficient implementation of rendering algorithms, such as, for example, ray-marching with sphere tracing \cite{hart1996sphere} and provides a modeling parameter for heterogeneous object modeling \cite{biswas2004heterogeneous}. On the other hand, it has a more limited set of modeling primitives and operations. 

\paragraph{Periodic functions}
Function Representation and SDF-based modeling have become increasingly popular in recent years for the purpose of modeling tiled and repeated patterns. This is used for modeling microstructures, which have proven useful in 3D printing among others \cite{pasko2011procedural}. Our system implements different operators for repeating (unit cell) patterns such as the triangle wave function and the sawtooth wave function. They can be combined with simple primitives (e.g., a sphere or a torus) to generate repetitive patterns as shown in \cite{pasko2011procedural}. Although the sawtooth or triangle wave functions are not differentiable at some points, it is also possible to replace them by their truncated Fourier series. We also provide implementations of several examples of triply periodic minimal surfaces such as gyroids and lidinoids, which are becoming extremely popular for modeling biological structures, among others \cite{han2018}.

\begin{figure}[htbp]
\begin{center}
 \includegraphics[width=0.3\textwidth]{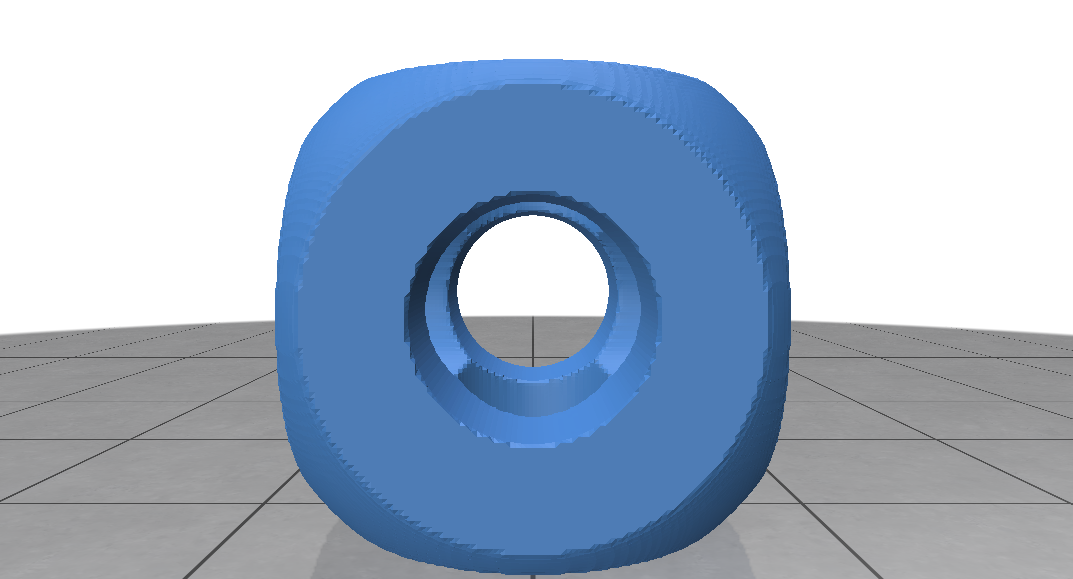}
 \includegraphics[width=0.3\textwidth]{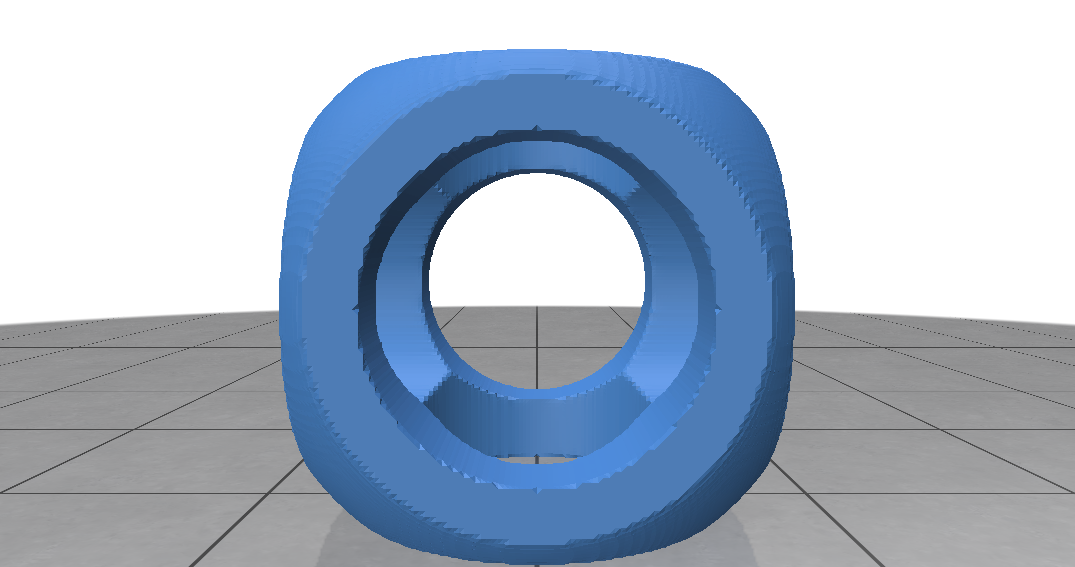}
 \includegraphics[width=0.3\textwidth]{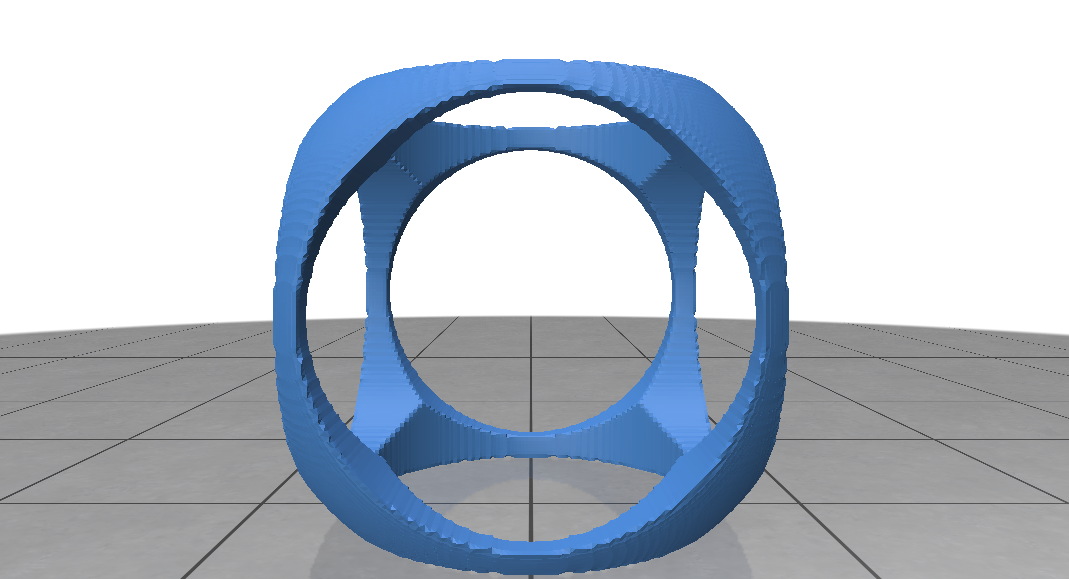}
\end{center}
\caption{Different shapes obtained by varying the radius of the cylinders (from left to right, $r=0.35,0.5,0.65$).}\label{fig:simple_object_r}
\end{figure}

\paragraph{Parametric modeling}
When modeling shapes for Computer Aided Design (CAD) applications, regular primitives (such as cuboids, spheres, cylinders, and others) are combined with Boolean operations (union, intersection, subtraction) and other operations (sweep, chamfer, and others). These primitives and operations can be parametrized, giving rise to parametric modeling \cite{Fayolle2008}. For example, a sphere can be parametrized by its center and radius (see the call to $sphere()$ in the model shown in Fig.\,\ref{fig:simple_object}). Varying these parameters allows one to explore different shapes in the modeling space. Figure \ref{fig:simple_object_r} shows, for example, different shapes obtained from the model in Fig.\,\ref{fig:simple_object} by varying the radius of the cylinders. 
If $f$ is the function defining a given solid, we indicate the dependency on shape parameters $\p \in \R^n$ by $f(\cdot; \p)$. The surface corresponding to a given set of parameters $\p$ is given by $\{\x\in\R^3: f(\x; \p)=0\}$. For a given $\x$, one can interpret $f(\x; \p)$ as a function of $\p$ and compute the derivatives with respect to these parameters. The computations are similar to the case of spatial derivatives. This allows us to fit a parametric model to a given data-set, such as, for example, a set of points $\{\x_1, \ldots, \x_N\}$ sampled on the surface of an object. This can be done by minimizing the loss function
\begin{equation}\label{eq:L2loss}
E(\p) = \frac{1}{N} \sum_{i=1}^{N} f(\x_i; \p)^2
\end{equation}
by Stochastic Gradient Descent (SGD) \cite{robbinsMonro1951}. Note that computing derivatives with respect to spatial variables $\x$, or shape parameters $\p$ are separate things. Even if a function is not differentiable with respect to the spatial variables at a given point, it may still be differentiable at that point with respect to its shape parameters.

\paragraph{Derivatives computation}
As described above, the core library is made of geometric primitives and operations written in pure Python. The derivatives can be computed with respect to the spatial parameters ($\x$ above) or with respect to the shape parameters ($\p$ above). We rely on PyTorch to create a computational graph for a given function and to compute the derivatives by reverse mode automatic differentiation. This is done by wrapping the corresponding variable in a Torch tensor, set to require derivatives computation. See, for example, the function in Fig.\ref{fig:grad} for computing $\nabla f(\x)$, the gradient of a function $f$ with respect to its argument $x$. 
\begin{figure}[htbp]
\begin{lstlisting}[language=Python,basicstyle=\small]
def compute_grad(f, x):
    if not torch.is_tensor(x):
        x = torch.tensor(x)
    x.requires_grad = True
    y = f(x)
    g = autograd.grad(y, [x],
                      grad_outputs=torch.ones_like(y),
                      create_graph=True)[0]
    return g
\end{lstlisting}
\caption{Computation of $\nabla f(\x)$.}\label{fig:grad}
\end{figure}
Computing the gradient of a model is useful w.r.t. both its spatial parameters or its shape parameters (for shape optimization). For the latter, usually one considers the gradient of a loss function, such as (\ref{eq:L2loss}), with respect to the shape parameters $\p$ (the spatial parameters $\x$ are fixed). We also provide functions for computing the divergence, the Laplacian (the divergence of the gradient), the $p$-Laplacian and the curvatures (principal, mean and Gaussian). These operators are mostly useful for the spatial derivatives only. The code for computing the divergence of a vector field is shown, for example, in Fig.\,\ref{fig:div}. 
\begin{figure}[htbp]
\begin{lstlisting}[language=Python,basicstyle=\small]
def compute_div(v, x):
    div = 0.0
    if not torch.is_tensor(x):
        x = torch.tensor(x)
    x.requires_grad = True
    y = v(x)
    for i in range(y.shape[-1]):
        div += autograd.grad(y[..., i],
                    x,
                    grad_outputs=torch.ones_like(y[..., i]),
                    create_graph=True)[0][..., i:i + 1]
    return div
\end{lstlisting}
\caption{Computation of the divergence of a vector field $\y=\v(\x)$}\label{fig:div}
\end{figure}
The computation of the gradient and the divergence are the only necessary ingredients to compute the different surface curvatures, the Laplacian ($\Delta f$) and the differential operators traditionally used in geometric modeling. 

\paragraph{Additional functionalities}
The rest of the framework contains common functionalities for performing input/output of standard mesh and point-cloud file formats, an implementation of the Marching Cubes algorithm \cite{lorensen1987marching} for rendering the zero level-set (surface) of a given model, functions for computing the different curvatures of a model, different normalization schemes, and a reinitialization algorithm, which is used to compute the signed distance function to the zero level-set of the given function. Finally, we propose implementations of different algorithms for fitting shape parameters according to different constraints, which are useful for inverse geometric modeling.

\section{Examples of application}
We describe in this section several possible applications: Automatic computation of the surface curvatures; Normalization schemes for distance approximation; Redistancing and shape fitting via a combination of heuristics and gradient descent.  

\subsection{Curvatures computation}
The principal curvatures of a surface are higher-order differential surface quantities that measure at a given point how the surface bends in a given direction. They are useful for shape interrogation. In our framework, a surface $\partial S$ is obtained from the zero level-set of a given function $\partial S = \{\x\in\R^3: f(\x)=0\}$. Computation of the curvatures involves the second order derivatives of $f$, which are obtained by automatic differentiation from the program specification of $f$. Unlike most existing approaches, we do not rely on any approximation such as, for example, approximating the surface by fitting a polynomial or a spline.  

\paragraph{Mean curvature}
For a surface point $\x$, the mean curvature is given by
\begin{equation}\label{eq:mean_curv}
H = -\frac{1}{2} \dive(\n), \quad \mathrm{where} \quad \n = \frac{\nabla f}{| \nabla f |}
\end{equation}
where $\dive$ is the divergence operator and $\n$ is the (inward) unit normal to the surface at the point $\x$, obtained from the normalized gradient of the defining function $f$. 

\paragraph{Gaussian curvature}
We use the expression of the Gaussian curvature given in \cite[Theorem 4.1]{goldman2005curvature} 
\begin{equation}\label{eq:gauss_curv}
K = \frac{\nabla f \cdot \mathrm{Hessian}^*(f) \cdot \nabla f^T}{|\nabla f|^4}
\end{equation}
where $\mathrm{Hessian}(f)$ is the Hessian matrix of $f$, $M^*$ denotes the adjoint of matrix $M$, and $M^T$ denotes the transpose of matrix $M$. 

\paragraph{Principal curvatures}
Finally, the principal curvatures $\kappa_{\min}$ and $\kappa_{\max}$ can be computed from the mean and Gaussian curvatures using 
$$
H = \frac{1}{2}(\kappa_{\min} + \kappa_{\max}) \qquad K = \kappa_{\min}\,\kappa_{\max}.
$$
Namely, we have
$$
\kappa_{\min} = H - \sqrt{H^2 - K} \qquad \kappa_{\max} = H + \sqrt{H^2 - K}.
$$

\begin{figure}[htbp]
\begin{center}
 \includegraphics[width=0.45\textwidth]{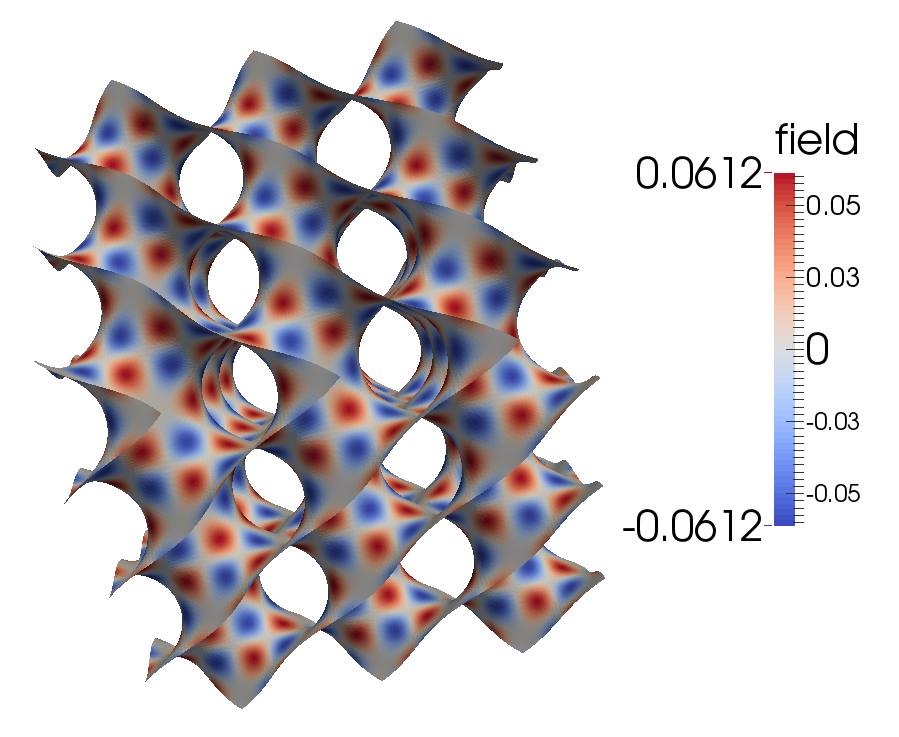}
 \includegraphics[width=0.45\textwidth]{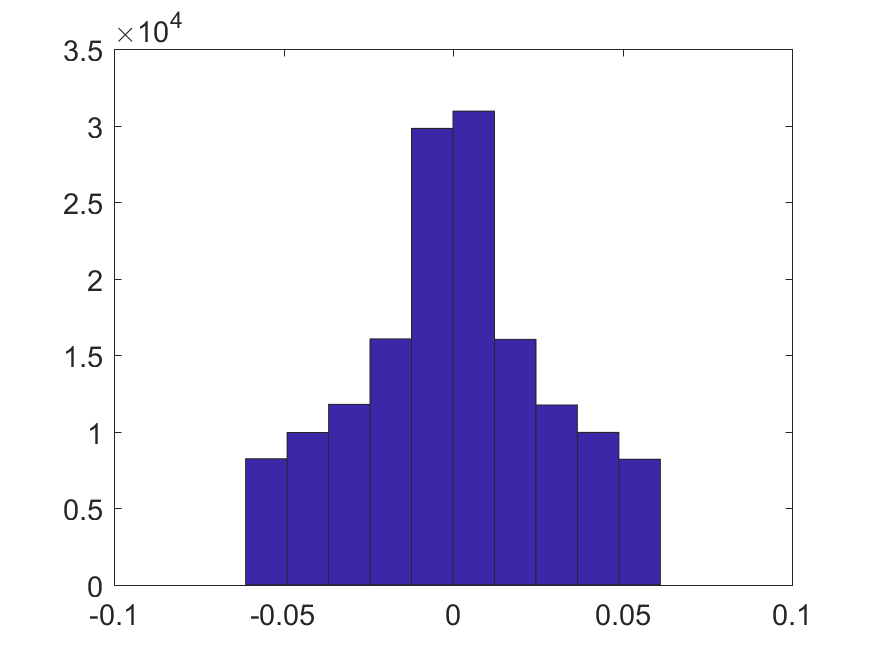}
\end{center}
\caption{Left: Mean curvature on a minimal surface (the Schwarz D minimal surface). Right: Distribution of the mean curvature values. }\label{fig:minimal_surface_curv}
\end{figure}

\paragraph{Example} 
The Schwarz D minimal surface is given by
\begin{align*}
& \sin(x)\sin(y)\sin(z) + \sin(x)\cos(y)\cos(z) \\
& + \cos(x)\sin(y)\cos(z) + \cos(x)\cos(y)\sin(z) = 0. 
\end{align*}
See Fig.\,\ref{fig:minimal_surface_curv}.
The surface (zero level-set) is computed by the Marching Cubes algorithm. The mean curvature is then computed at each vertex of the triangle mesh obtained from the Marching Cubes algorithm by (\ref{eq:mean_curv}). In the case of the minimal surface, we know that $H = 0$. We can experimentally verify that the computed values for the mean curvature are close to zero in Fig.\,\ref{fig:minimal_surface_curv}. This is also confirmed by looking at the distribution of the mean curvature values in Fig.\,\ref{fig:minimal_surface_curv}, right. 

\subsection{Normalization}
Normalization is a technique used to compute an approximate distance from a point to implicitly defined surfaces (curves in the two-dimensional case) \cite{Rvachev_book74,Taubin_tog94}. Assume that the boundary $\partial S$ is the zero level-set of a function $f(\x)$, then the distance $d(\x)$ can be estimated by \cite{Rvachev_book74}
\begin{equation}\label{omega1}
\omega_1[f](\x)=\frac{f(\x)}{\sqrt{f(\x)^2+|\nabla f(\x)|^2}}.
\end{equation}
The normalization $\omega_1[f]$ satisfies the boundary conditions $\omega_1[f](\x)=0$ and $\partial \omega_1[f]/\partial\n=1$ on $\partial S$

A normalization to the order $k$ is obtained recursively by 
\begin{equation}
\omega_k[f](\x) = \omega_{k-1}[f](\x) - \frac{1}{k!} \omega_1[f]^k \frac{\partial^k \omega_{k-1}[f]}{\partial \n^k},  
\end{equation} 
and satisfies the boundary conditions  
\begin{eqnarray}
&&\omega_k[f]=0\quad\mbox{and}\quad\partial \omega_k[f]/\partial\n=1
\quad\mbox{on}\quad\partial S,
\\
\nonumber
&&\partial^{\,l}\!\omega_k[f]/\partial\n^l=0\quad\mbox{on}\quad\partial S,
\quad l=2,3,\dots,k.
\end{eqnarray}

An alternative normalization scheme is given by \cite{Taubin_tog94}
\begin{equation}\label{delta1}
\!\!\delta_1[f](\x)=f(\x)/|\nabla f(x)|.
\end{equation}

\begin{figure}[htbp]
\begin{center}
 \includegraphics[height=3cm]{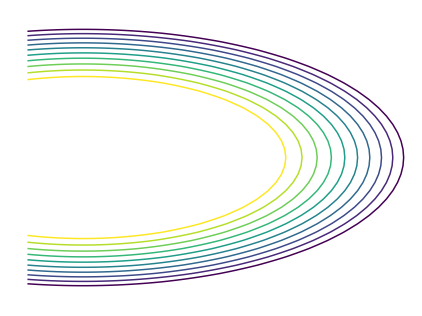}
 \includegraphics[height=3cm]{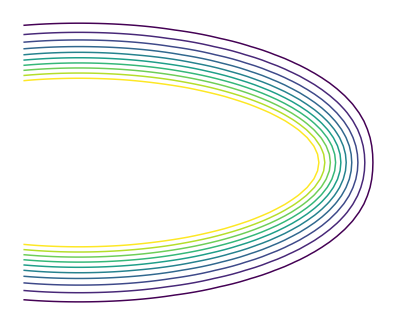}
 \includegraphics[height=3cm]{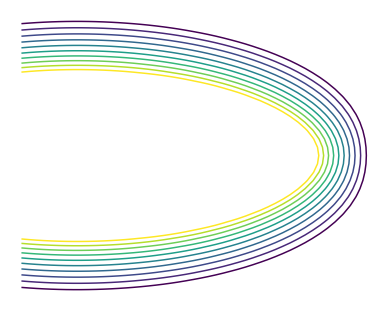}
\end{center}
\caption{Left: Level-sets (from -0.5 to 0.5) of an ellipse with semi-major and semi-minor axes of 5 and 2. Middle: Rvachev normalization $w_1$. Right: Taubin normalization $\delta_1$}\label{fig:normalization}
\end{figure}

All these normalization schemes are based on the computation of the spatial derivatives of the model $f(\cdot)$. Although $\omega_1[f]$ and $\delta_1[f]$ satisfy appropriate boundary conditions and provide good approximations of the distance function near the surface, their behavior is much less accurate away from the boundary. The normalization of an ellipse is shown in Fig.\,\ref{fig:normalization}.  

\subsection{Distance Function Estimation}
\label{sec:redistancing}
We consider the problem of reinitialization, also known as redistancing, of implicit surfaces. The problem is as follows: Given $f:\R^3\rightarrow\R$, whose zero level-set defines a surface $\partial S$, we want to compute the signed distance function (SDF) $d:\R^3\rightarrow\R$ coinciding with $f$ on $\partial S$. The function $f$ is built here with the approach described in Section~\ref{sec:system} and appears implicitly in the definition of $\partial S$.

For computing the SDF $d(\cdot)$, we use the approach introduced in \cite{fayolle2021signed}: It uses as an ansatz for $d(\x)$ the function $\mathrm{sign}(f(\x))h(\x;\btheta)$, where $h(\cdot;\btheta)$ is a fully connected deep neural network, and $\mathrm{sign}(\cdot)$ is the sign function. Multiplying the final layer of the deep neural network by $\mathrm{sign}(f(\x))$ guarantees that the zero level-set of $f$ is preserved. The parameters of $h(\cdot;\btheta)$ are obtained by minimizing the loss function 
\begin{equation}\label{loss_eik}
L(\btheta) = \mathbb{E}_{\x \sim D} \left(\left|\nabla_{\x} d(\x;\btheta)\right|-1\right)^2, 
\end{equation}
where $\mathbb{E}$ is the expectation, and $D$ is the uniform distribution over the given computational domain. Automatic differentiation is used to compute the spatial derivatives, as well as the derivatives w.r.t. the parameters $\btheta$. 

\begin{figure}[htbp]
\begin{center}
 \includegraphics[width=0.45\textwidth]{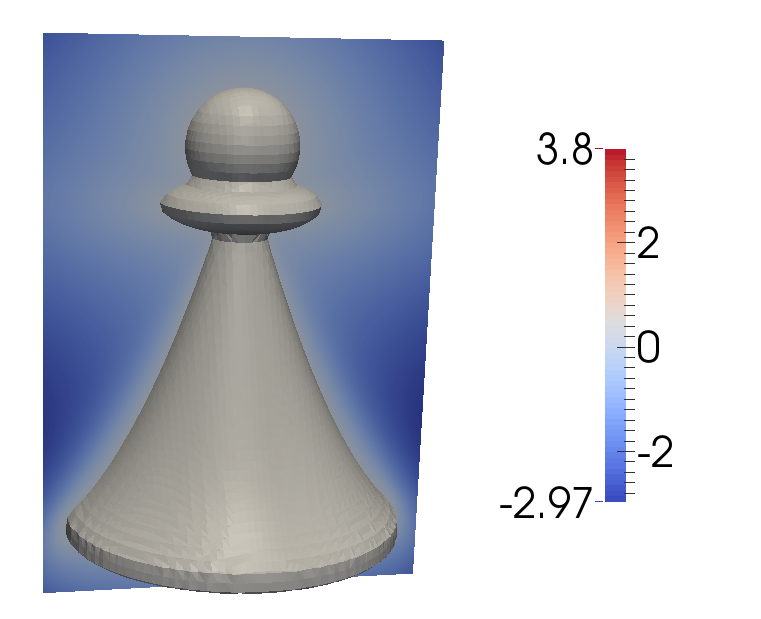}
 \includegraphics[width=0.45\textwidth]{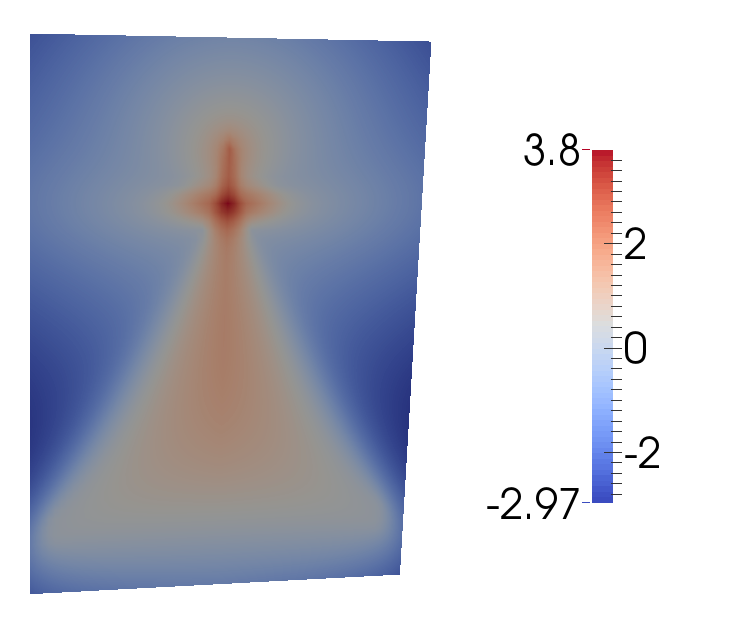}
\end{center}
\caption{Left: Zero level-set of $f$ and filled contour plot on a slice. Right: Filled contour plot on a slice without the surface.}\label{fig:pawn}
\end{figure}

Figure\,\ref{fig:pawn} shows a chess pawn modeled implicitly, using simple primitives (polynomials) and Boolean operations. The image on the left illustrates the surface $\{\x\in\R^3: f(\x)=0\}$ and the filled contour plot of $f$ on a slice. The image on the right shows the filled contour plot of $f$ without the surface to illustrate the field behavior in the interior of the object as well. 

\begin{figure}[htbp]
\begin{center}
 \includegraphics[width=0.45\textwidth]{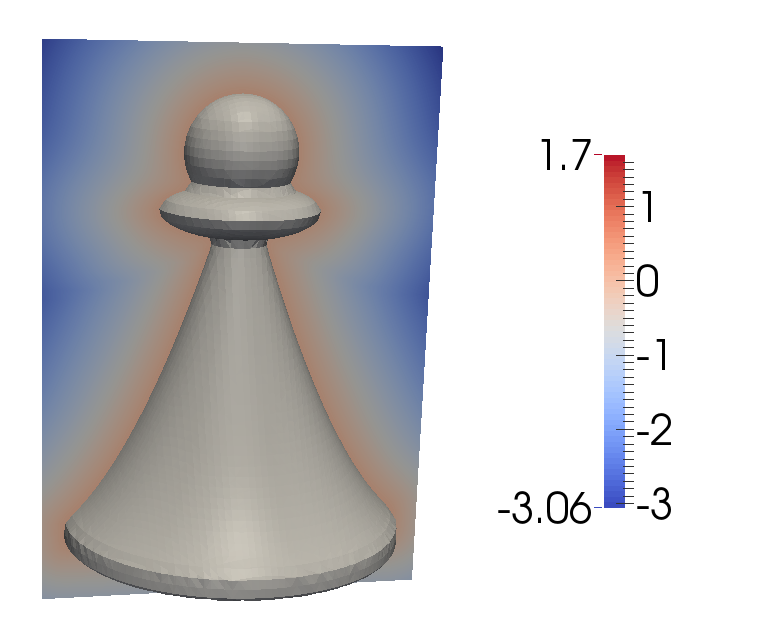}
 \includegraphics[width=0.45\textwidth]{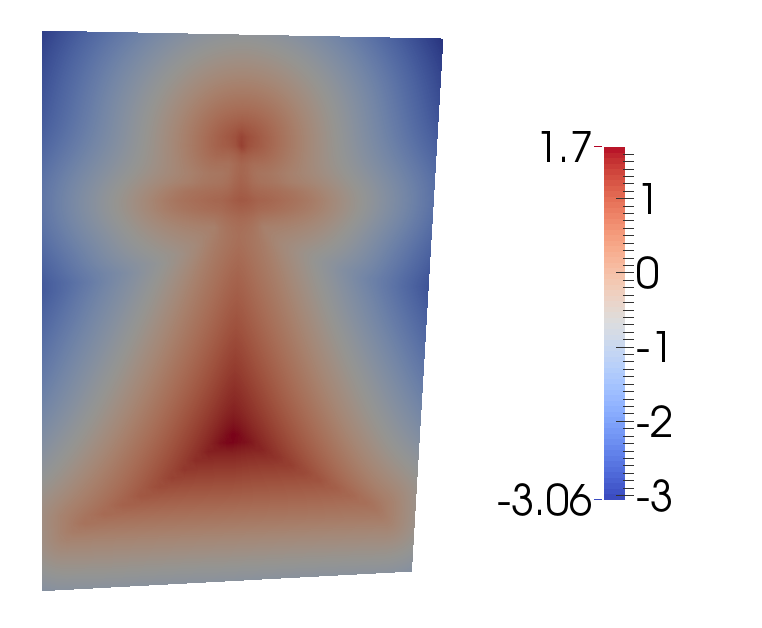}
\end{center}
\caption{Left: Zero level-set of $d(\x) = \mathrm{sign}(f(\x)) h(\x;\widetilde{\btheta})$ and filled contour plot on a slice. Right: Filled contour plot on a slice without the surface.}\label{fig:pawn_dist}
\end{figure}

Figure\,\ref{fig:pawn_dist} shows the result obtained for the distance function $d(\x)$. Compare the filled contour plots with Fig.\,\ref{fig:pawn} and notice that it delivers a much improved approximation of the distance function to the surface.  

\subsection{Parametric Shape Fitting}
\label{sec:fitting}
We show how the proposed framework can be used to fit parametric models to data and used in inverse geometric modeling \cite{Fayolle2008}. Let $f(\x;\p)$ be a parametric model, where $\p$ are the parameters controlling the shape of the object. For a given set of parameters $\p$, the corresponding solid is given by the set $\{\x\in\R^3: f(\x;\p)\geq 0\}$. 

\begin{figure}[htbp]
\begin{center}
 \includegraphics[height=3.5cm]{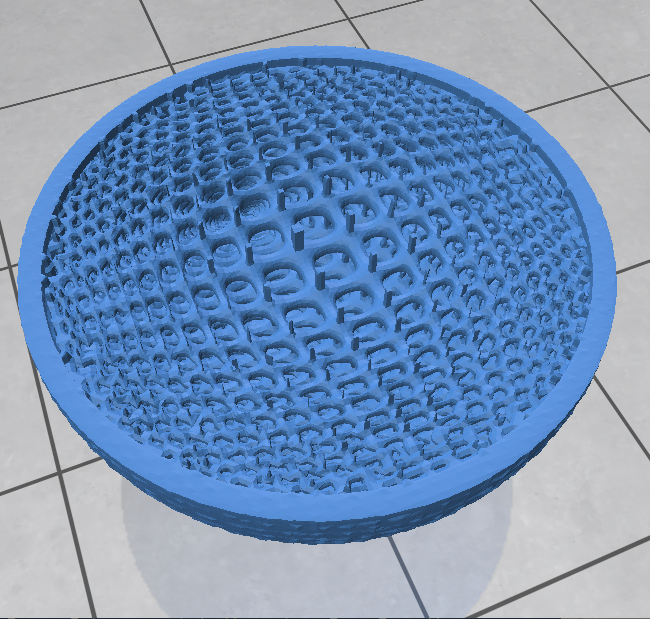}
 \includegraphics[height=3.5cm]{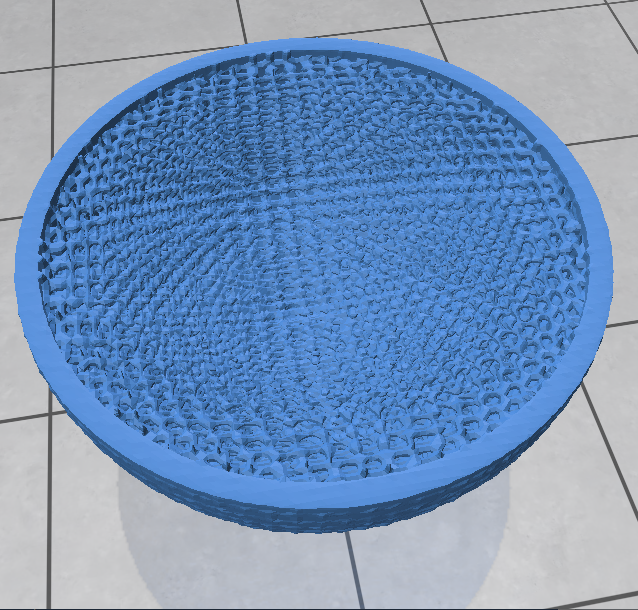}
 \includegraphics[height=3.5cm]{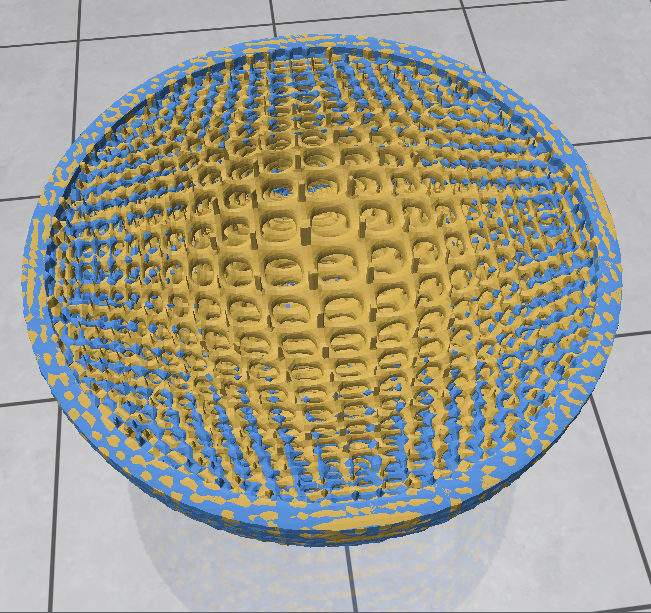}
\end{center}
\caption{Reconstructed object from fitting a parametric model. Left: Regularized evolution + SGD is used. Middle: SGD alone. Right: Both models are shown together. Yellow: Regularized evolution + SGD; Blue: SGD alone.}\label{fig:fitting_meshes}
\end{figure}

Let $\{\x_1,\ldots,\x_N\} \subset \R^3$ be the input point-cloud, i.e. the set of 3D points acquired on the surface of the object. The goal is to find the parameters $\p$ such that the shape corresponding to $f$ fits the input point-cloud. Each parametric model comes with a set of shape parameters that controls the appearance of the model. For example, the parameters of the model shown in Fig.\,\ref{fig:fitting_meshes} correspond to the frequency and scaling of the rods. 

In order to fit the model parameters to a given dataset, we define the error of fit by 
\begin{equation}\label{eq:E(p)}
E(\p) = \frac{1}{N}\sum_{i=1}^{N} f(\x_i; \p)^2
\end{equation}
where $\p$ are the model parameters and $\x_i\in\R^3$ are the points from the input point-cloud (or the vertices of a given triangle mesh). $E(\p)$ has to be minimized for $\p$.  

The minimization is done by a combination of a heuristic method and stochastic gradient descent. We use a variant of regularized evolution \cite{real2019regularized} as a heuristic method to find parameters approaching a global minimum of (\ref{eq:E(p)}). Regularized evolution is a type of evolutionary algorithm in which a parent, selected at each iteration by tournament selection, is then mutated and inserted back into the population. The oldest creature of the population is removed at each iteration. We use then Stochastic Gradient Descent (SGD) \cite{robbinsMonro1951} to improve the solution
$$
\p_{i+1} \leftarrow \p_{i} - \nabla_{\p} \widetilde{E}(\p_i), 
\quad \mathrm{where} \quad 
\widetilde{E} = \frac{1}{|I|}\sum_{i\in I} f(\x_i; \p)^2, 
$$
where $I$ is a random subset of $\{1,\ldots,N\}$, the initial parameters $p_0$ are obtained from regularized evolution, and the derivatives of the model with respect to the shape parameters $\p$ are computed by automatic differentiation. 

We use as an example a parametric microstructure built by repeating rods (see Fig.\,\ref{fig:fitting_meshes}). The parameters of the model control the frequency of the rods and their sizes. Given as input a point-cloud made of $40K$ points, we fit the parametric model to the input point-cloud. We use $10K$ iterations of regularized evolution with a population size of $100$ and a sample size of $10$, followed by $100$ iterations of SGD.
\begin{figure}[htbp]
\begin{center}
 \includegraphics[height=4cm]{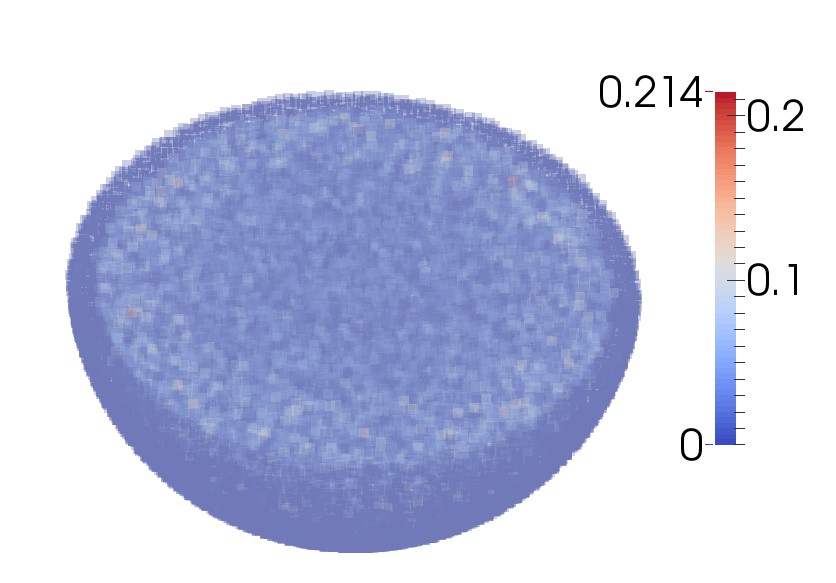}
 \includegraphics[height=4cm]{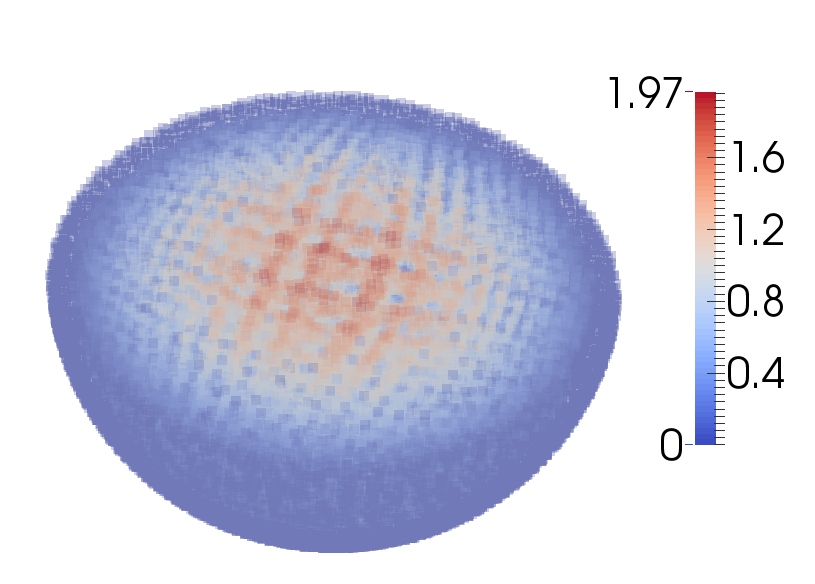}
\end{center}
\caption{Point-wise absolute error for the fitted model. Left: Regularized evolution + SGD is used. Right: SGD alone. Notice the large errors committed for the model fitted with SGD alone (right).}\label{fig:fitting_errors}
\end{figure}
Figures\, \ref{fig:fitting_errors} and \ref{fig:fitting_meshes} show the results obtained by the approaches described above. The absolute value error ($|f(\x_i;\tilde{\p})|$ corresponding to the fitted parameters $\tilde{\p}$) is shown at each point of the input point-cloud for the models fitted by SGD alone in Fig.\,\ref{fig:fitting_errors}, right image, and by the combination of regularized evolution and SGD described above in Fig.\,\ref{fig:fitting_errors}, left image. Note the larger errors committed for the model fitted with SGD alone, compared to the model fitted by regularized evolution with SGD. 

The surfaces corresponding to the models fitted by SGD alone (middle image) and regularized evolution + SGD (left image) are shown in Fig.\,\ref{fig:fitting_meshes}. The rightmost image shows both surfaces together with the result of regularized evolution + SGD in yellow, and the result of SGD alone in blue color. 

\section{Conclusion}
We described a framework for differentiable geometric modeling based on the Function Representation. The proposed system is built on top of automatic differentiation libraries, allowing to easily compute and obtain the derivatives of a geometric object w.r.t. the spatial, as well as the shape parameters. We demonstrated several possible applications of this framework: Curvatures computation, normalization, signed distance estimation, and parametric model fitting. The code is open-source and we hope that it can be further extended and used in geometric pipelines. 

\small
\bibliographystyle{plain}
\bibliography{paper}

\end{document}